# On Synergy effect of Ohkawa Current Drive of Electron Cyclotron Waves and Lower Hybrid Current Drive: A New Mechanism

P. W. Zheng,[1,2] X. Y. Gong,[1,*] X. Q. Lu,[1] L. H. He,[1] J. J. Cao,[1] Q. H. Huang,[1] S. Deng,[1] J. F. Lin,[1] Y. J. Zhong[3]

1. School of Nuclear Science and Technology, University of South China, Hengyang, Hunan 421001, P. R. China
2. School of Environmental and Safety Engineering, University of South China, Hengyang, Hunan 421001, P. R. China
3. School of Mathematics and Physics, University of South China, Hengyang, Hunan 421001, P. R. China

A new synergy mechanism between Ohkawa current drive (OKCD) of electron cyclotron (EC) waves and lower hybrid current drive (LHCD) is discovered and discussed. And the methodology to achieve this synergy effect is also introduced. Improvement of OKCD efficiency can be achieved up to a factor of ~ 2.5 in far off-axis radial region ($\rho > 0.6$) of tokamak plasmas. Making EC wave heating the electrons of co-$I_p$ direction and LH wave heating the electrons of counter-$I_p$ direction, the mechanism of this new synergy effect comes from the results of electron trapping and detrapping processes. The OKCD makes the low speed barely passing electrons to be trapped (trapping process), the LHCD pulls some of the high speed barely trapped electrons out of the trapped region in velocity space (detrapping process) and accelerates the detrapped electrons to a higher speed.

*Introduction.*—The synergy effect of electron cyclotron (EC) waves and lower hybrid (LH) waves is of great interest in the area of non-inductive current drive in tokamaks. Although with limited control capabilities, LH waves are believed to provide the highest current drive efficiency. EC waves have lower efficiency comparing to LH waves, but they can drive highly localized currents. Previous study on the Tore Supre tokamak demonstrated both in theory and experiment that strong synergy effect existed [1] for the two kinds of radio-frequency (RF) waves, improving (up to a factor ~4) the EC current drive (ECCD) efficiency in plasmas sustained by LH current drive (LHCD). The synergy effect between ECCD and LHCD was found in the inner half region of the Tore Supre tokamak (normalized radius $\rho \leq 0.4$). The mechanism of the synergy effect is that EC waves push low-velocity electrons out of the Maxwellian bulk, and LH waves accelerate them to high parallel velocities.

The interaction mechanism of LH waves with electrons is parallel velocity diffusion associated with the Landau damping [2-3]. Hence, the LHCD is poorly insensitive to electron trapping, inducing electron detrapping rather than trapping, and possibly pulling some barely trapped electrons out of trapped region in velocity space. Ohkawa current originates from selective electrons trapping [3-4]. Perpendicular diffusion of EC waves causes the barely resonant passing electrons to become trapped, which induces a net reverse parallel current. This reverse current is known as the Ohkawa current. It is generally believed to be small, but it reduces the current drive efficiency greatly for off-axis ECCD [5-6], which is originated from the Fisch-Boozer mechanism [7]. However, recent theoretical study shows that Ohkawa current drive (OKCD) of EC waves has the potential capability in driving an effective localized current in the outer half region of tokamak plasma with a large inverse aspect ratio or in the radial position where the local inverse aspect ratio is large enough [8-9]. So that synergy effect between OKCD and LHCD possibly be exist in far off-axis region of a tokamak. In this letter, we demonstrated theoretically that the strong synergy effect between OKCD and LHCD can exist in the outer half region (far off-axis) in a low inverse aspect ratio tokamak.

*Simulation details.*—The GENARAY [10] /CQL3D [11] package suite codes and the EAST H-mode discharge *#013606* are used in simulation. The EAST is a medium sized tokamak with major radius $R_p = 1.85m$, minor radius $a = 0.465m$, inverse aspect ratio of $\varepsilon = a/R_p \sim 0.25$. The configuration of EAST, the electron density, the electron temperature, and the safety factor profiles are shown in FIG. 1. The equilibrium magnetic field is from EFIT [12] reconstruction with a plasma current $I_p = 0.8MA$. Central electron temperature in this discharge $T_{e0} = 7.28 keV$, central electron density $n_{e0} = 2.5 \times 10^{19} m^{-3}$, central magnetic field $B_{T0} = -2.2T$, and effective ion charge $Z_{eff} = 1.5$. One of the LHCD system in EAST operates with frequencies of 2.45GHz, the power spectra peaked at $n_{//} = 2.1$, initial spectral width $\Delta n_{//} = 0.26$ [13], and the LH wave is launched from low field side (LFS) mid-plane. And the EC wave is assumed with a frequency of 105GHz (*X2*-mode) in order to generate an effective OKCD in the outer half region of the EAST tokamak. For the purpose of making the current direction of effective OKCD have the same direction of LHCD, toroidal launch angle of the EC wave is fixed at $\alpha = 160°$, which corresponds to initial parallel refractive index of -0.342. In other words, the EC wave propagates in the opposite toroidal direction against the LH wave. The LH wave heats electrons of counter-$I_p$ direction, whereas the EC wave heats electrons of co-$I_p$ direction. The radial power deposition of the EC wave is varied by changing poloidal incident angle $\beta$ of the EC wave to make the EC and the LH powers are overlapped at the same radial location. The EC wave is also launched from the LFS mid-plane. The GENRAY ray-tracing code is used to obtain wave parameters and power damping of the two RF waves. The EC and the LH wave beams propagation are described by tracing 48 rays and

50 rays per beam, respectively. Summing up all the contributions of the GENRAY code to the quasi-linear diffusion coefficient, the calculations are performed with the CQL3D Fokker-Planck quasi-linear simulation code.

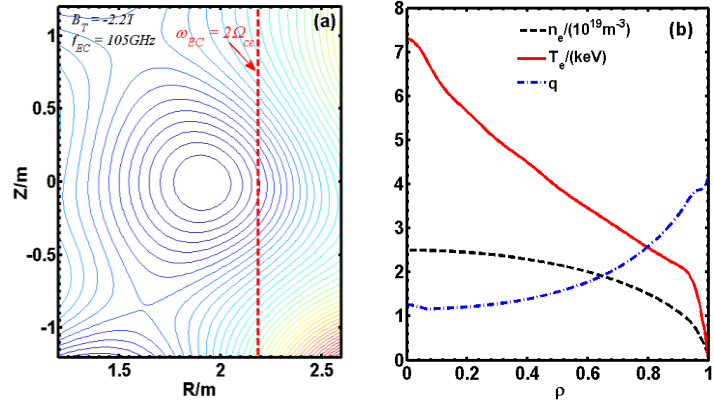

**FIG.1** (color online). (a) Flux surfaces of the plasma equilibrium of EAST discharge *#013606*, the resonance domain located at *R* = 218.52cm for second harmonic resonance of the EC wave with frequency 105GHz, and (b) profile of the electron density, temperature and safety factor.

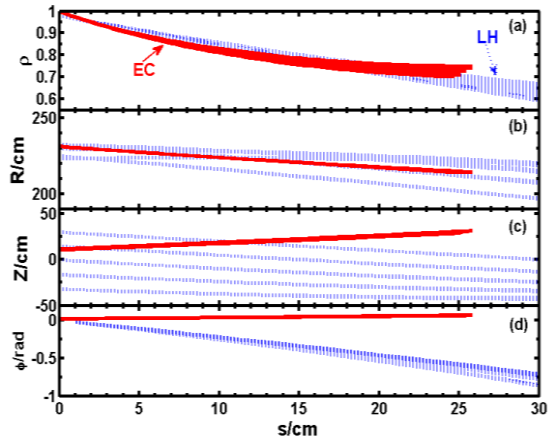

**FIG. 2** (color online). Ray trajectories of the EC (solid red lines) and the LH (dotted blue lines) waves calculated by the GENRAY code, (a) normalized toroidal flux surface labels $\rho$ versus poloidal distance s along the all rays, (b)-(d) the *R*, *Z* and the toroidal angle of the rays versus poloidal distance *s*.

*Simulation results and analysis.*—FIG. 2 shows the ray trajectories of the EC and LH waves calculated by the GENRAY code. As shown in FIG. 2(d), the EC and the LH waves propagate along different toroidal directions of the tokamak, heating the passing electrons with different directions along the magnetic field lines of tokamak. The FIG. 2(a)-(c) shows that the ray trajectories of the EC and LH waves are overlapped near the LFS mid-plane, so that the power of the two waves is likely to be deposited in the same radial region.

The desired radial location $\rho_{EC}$ of the EC driving current is varied by changing the poloidal launch angle $\beta$, because the toroidal launch angle is fixed at $\alpha = 160^0$ for an effective OKCD. The power deposition and the driven current profiles of the two waves are shown in FIG. 3. FIG. 3(a) shows that the power of the two waves is deposited in the radial range of $\rho = 0.70$ ~ 0.85 simultaneously. Following the references [1,14], the synergy effect is quantified by synergy factor $F_{syn} = \Delta I/I_{EC}$, where $\Delta I = I_{LH+EC} - I_{LH}$, $I_{LH+EC}$ is the current driven by the simultaneous use of the two waves LH+EC, $I_{LH}$ and $I_{EC}$ are total driven current by the single LH and the single EC, respectively. FIG. 3(b) shows the radial driven current profiles for this case. The black solid line represents the profile of the contribution of the two waves simultaneous. The green dash-dotted and blue dotted lines express the profiles of the LH alone and the EC alone, respectively. And the red solid line indicates the synergy driven current profile, which defined as $j_{EC+LH}(\rho) - j_{EC}(\rho) - j_{LH}(\rho)$. In this case, the power of the EC and the LH waves is 1.0MW, respectively. The current driven by the simultaneous use of the EC and the LH waves is $I_{LH+EC} = 376.3$kA, the current driven by the EC wave alone is $I_{EC} = 21.9$kA, and $I_{LH} = 338.0$kA. The synergy factor $F_{syn}$ for this case is 1.75, which means a moderate synergy effect of the two waves.

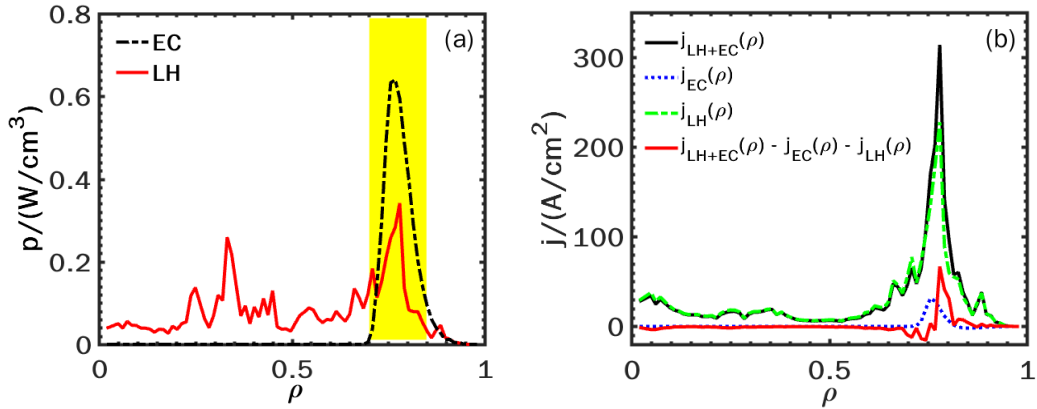

**FIG. 3** (color online). (a) Profiles of the EC and the LH power absorption versus normalized toroidal flux surface labels $\rho$. (b) Radial driven current profiles versus $\rho$ for LH+EC, EC alone, LH alone, and net synergy current. The peak value of the net synergy current locates at $\rho = 0.78$. Results are calculated by the GENRAY/CQL3D code for 1MW of the EC wave and 1MW of the LH wave.

As shown in FIG. 3 (b), the synergy driven current profile appears to be shifted slightly to the right compared to the current profile of the EC driven alone, while aligned with the current profile of the LH driven alone. It means that the EC and the LH power have a different impact on the synergy effect between OKCD and LHCD. Scan of the $P_{LH}/P_{EC}$ and $P_{EC}/P_{LH}$ on the synergy factors are performed for the EC power deposited at three different radial locations. FIG. 4 shows that increase of both the EC and the LH power can improve the synergy factors, but the LH power makes a bigger contribution to the $F_{syn}$ compared to the EC power. FIG. 4 (a) shows that the synergy factor increases with the increasing of the $P_{LH}/P_{EC}$, ranging from ~1.5 for $P_{LH}$ = 0.5MW to ~2.5 for $P_{LH}$ = 5.0MW at the $\rho$ = 0.78. Varying the poloidal incident angle $\beta$ to make the EC power deposited in the radial region of $\rho$ ~ 0.67 - 0.78, the synergy factor can achieve a higher value (up to ~ 3.4). However, keeping $P_{LH}$ = 1.0 MW constant, the synergy factor increases slowly with the increase of the EC power. This indicates that the LH wave contributes more to the synergy effect under the same power level. The synergy factor is less sensitive to electron trapping effect due to the fact that the Ohkawa currents of EC waves are originated from this effect. This can be reflected from the results in FIG. 4. The synergy factors change moderately at the three different radial location for a constant ratio of $P_{LH}$ to $P_{EC}$.

FIG. 5 shows special profiles of driven current density versus normalized velocity $u/u_{norm}$ at the radial location $\rho$ = 0.78 for the same power level (1MW) of the EC and the LH waves, respectively. The green solid line represents Ohkawa current driven by the EC wave alone, the Ohkawa current is mainly driven by low velocity electrons with $u/u_{norm} \approx 0.12$. The blue dash-dotted line represents LHCD alone, which shows the vast majority of the total current is driven by the Landau damping in spite of a little amount of reverse Ohkawa current is driven by the LH wave at $u/u_{norm} \approx 0.205$. The red dash-dotted line expresses the combined current drive profile of the EC and the LH waves. And sum of the EC alone and the LH alone is marked with blue dash-dotted line. Comparing the red dash-dotted and the blue dash-dotted lines, the synergy effect at $\rho$ = 0.78 is mainly coming from the contribution of high energy electrons ($u/u_{norm} \geq 0.26$) which are accelerated by the LH wave in parallel direction. Because the synergy current profile (marked by the red dash-dotted line) in $u/u_{norm}$ space ranging of 0 ~ 0.16 is almost the same as the Ohkawa current profile (marked by the green solid line) driven by the EC wave. This synergy effect can not be ascribed to the following reasons, the EC wave pushes some barely passing electrons with low velocities into the trapped region, the

LH wave pulls these electrons out of the trapped region and accelerates them to higher velocities.

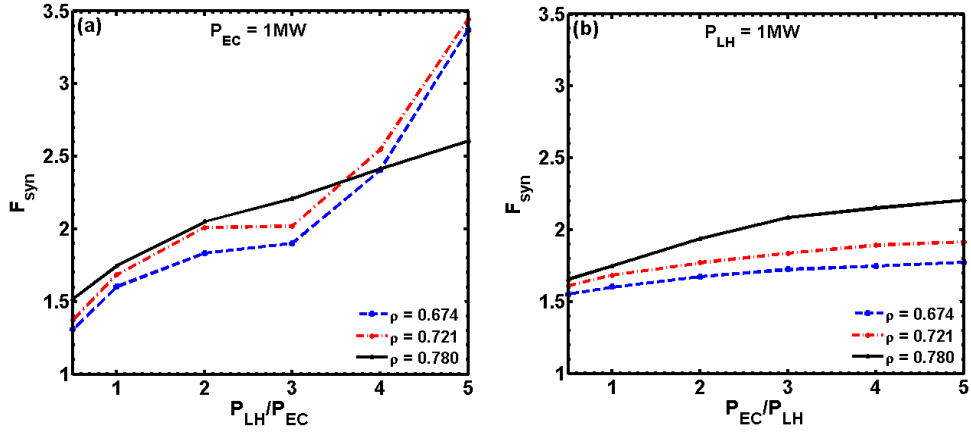

**FIG. 4** (color online). Synergy factors versus (a) $P_{LH}/P_{EC}$ and (b) $P_{EC}/P_{LH}$ for EC power deposited at three different radius locations $\rho$.

In order to reveal this new synergy mechanism between OKCD and LHCD, contours of quasi-linear diffusion strength due to RF($u^2 D_{uu}$) in velocity space are plotted separately both for the EC alone and the LH alone. Where $D_{uu}$ is the quasi-linear diffusion coefficient in the velocity space. The synergy effect can be interpreted as shown in the FIG. 6. The OKCD makes the barely passing electrons to be trapped (trapping process). The LHCD pulls some of the barely trapped electrons out of trapped region in velocity space (detrapping process) and accelerates these detrapped electrons to a higher speed. The trapping process induced by the EC wave is mainly acts on the barely passing electrons with low velocities ($u/u_{nom} \sim 0.12$), these electrons are locating just below the right trapped/passing boundary (TPB) in the velocity space. The detrapping process induced by the LH wave is mainly acts on the barely trapped electrons with high velocities ($u/u_{nom} \geq 0.26$) nearby the left TPB. Furthermore, as shown in FIG. 6, the quasi-linear diffusion strength due to the LH wave is much more stronger than the EC wave. Once these barely trapped electrons with higher velocities are pulled out of the trapped region, they are more easily to be accelerated by the LH wave to a higher parallel speed in the counter-$I_p$ direction, and subject to less collisional resistance, resulting in contributing more to the net synergy current of the two waves. It is for this reason that increase of both the EC and the LH power can improve the synergy factor, as shown in FIG. 4, while the increase in LH power has a greater impact on the synergy effect, making a larger synergy factor than the EC wave at the same power level.

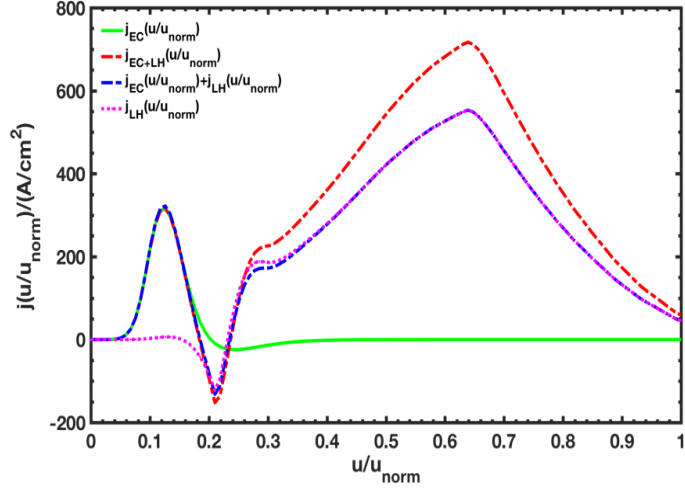

**FIG. 5** (color online). Profiles of driven current versus normalized velocity $u/u_{norm}$ at the radial location $\rho =$ 0.78 for 1MW EC power and 1MW LH power by the CQL3D code. The green solid line represents the EC alone, the pink dotted line indicates the LH alone, and the red dash-dotted line expresses EC+LH. The sum of the EC alone and LH alone are also marked with blue dash-dotted line.

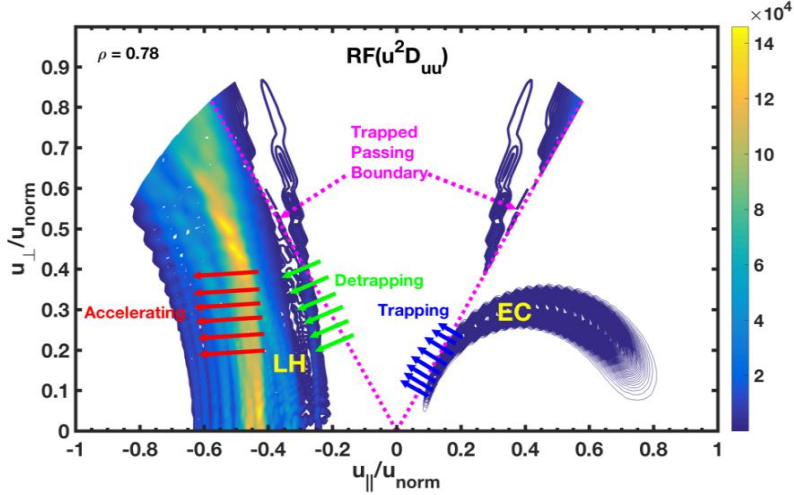

**FIG. 6** (color online). The physical picture of the synergy effects between OKCD and LHCD in velocity space. $P_{EC}$ = 1.0MW, $P_{LH}$ = 1.0MW, $\rho_{EC}$ = 0.78. Contours represent the LH and EC quasi-linear diffusion strength due to RF($u^2D_{uu}$) in velocity space with the perpendicular and parallel velocities normalized to $u_{norm}$, respectively. $u_{norm}$ corresponds to 300$keV$ kinetic energy of electron. The pink dotted lines are the TPBs.

*Conclusion and discussion.*—In conclusion, a new synergy mechanism between OKCD of EC waves and LHCD was discovered in far off-axis radial region. Strong synergy effect (up to a factor of $F_{syn}$ ~ 2.5) exists in the outer half region in the EAST tokamak with a low inverse aspect ratio $\varepsilon$ ~ 0.25. The methodology to achieve this synergy effect is introduced.

The major conditions include (i) EC waves and LH waves must heat passing electrons with different directions paralleling to the background magnetic field lines of tokamak plasmas, respectively; (ii) the EC parameters must be carefully selected to let the resonant region of EC waves just below the TPBs in velocity space, making the Ohkawa current drive mechanism dominant over the Fisch-Boozer current drive mechanism of EC waves; (iii) both the LH power and the EC power must be deposited simultaneously on the LFS mid-plane in the far off-axis radial region, and the profiles of the power deposition for the two waves also must be overlapped in radial region. The underlying mechanism of the synergy effect is revealed and discussed. The OKCD pushes some of the barely passing electrons with low velocities ($u/u_{nom}$ ~ 0.12) to be trapped (trapping process), the LHCD pulls some of the barely trapped electrons with high velocities ($u/u_{norm} \geq 0.26$) out of the trapped region in velocity space (detrapping process) and accelerates the detrapped electrons to a higher speed. Increase of both the EC and the LH power can improve the synergy factor. The synergy effect is more sensitive to the increase of LH power than EC power, making a larger synergy factor than the EC wave at the same power level. The new synergy effect is less sensitive to electron trapping. Therefore, this new synergy effect may be of useful application in far off-axis current drive for current profile control and suppression of some important MHD instabilities.

This work was supported by National Natural Science Foundation of China (No. 11405082, 11675073, 11375085), Hunan Provincial Natural Science Foundation of China (2018JJ2320), and the Construct Program of Fusion and Plasma Physics Innovation Team in Hunan Province (NHXTD03). The authors would like to thank Bob Harvey at CompX for providing GENRAY and CQL3D codes.

*gongxueyu@126.com
[1] G. Giruzzi, et al., Phys. Rev. Lett. **93**, 255002 (2004).
[2] N. J. Fisch, Phys. Rev. Lett. **41**, 873 (1978).
[3] N. J. Fisch, Rev. Mod. Phys. **59**, 175 (1986).
[4] T. Ohkawa, *General Atomics Report GA-A13847* (San Diego, CA: General Atomics, 1976). (http://library.psfc.mit.edu/catalog/online_pubs/tech_reports/GA-A13847.pdf)
[5] C. C. Petty, et al., Nucl. Fusion **42**, 1366 (2002).
[6] C. C. Petty, et al., Nucl. Fusion **43**, 700 (2003).
[7] N. J. Fisch and A. H. Boozer, Phys. Rev. Lett. **45**, 720 (1980).
[8] P. W. Zheng, et al., Nucl. Fusion **58**, 036010 (2018).
[9] P. W. Zheng, et al., Phys. Plasmas **25**, 072501 (2018).

[10] A.P. Smirnov and R.W. Harvey, *Report CompX-2000-01* (CompX, 2003). http://www.compxco.com/Genray_manual.pdf

[11] R.W. Harvey and M. G. McCoy, in *Proceedings of the IAEA Technical Committee Meeting on Numerical Modeling of Plasmas, Montreal, 1992* (IAEA, Vienna, 1993).

[12] L. Lao, et al., Nucl. Fusion **25** 1611 (1985).

[13] M. H. Li, et al., Phys. Plasmas **23**, 02512 (2016).

[14] R. J. Dumont, G. Giruzzi, and E. Barbato, Phys. Plasmas 7, 4972 (2000).